\documentclass[aps,pra,twocolumn,showpacs,superscriptaddress,shortbibliography]{revtex4-2}
\usepackage{graphicx} 
\usepackage{amsmath}
\usepackage{soul}
\usepackage{graphicx,epstopdf}
\usepackage{gensymb}
\usepackage[dvipsnames]{xcolor}
\usepackage{footnote}
\usepackage{physics}
\usepackage{multibib}

\newcommand{\be}{\begin{equation}}
	\newcommand{\ee}{\end{equation}}
\newcommand{\bea}{\begin{eqnarray}}
	\newcommand{\eea}{\end{eqnarray}}
\newcommand{\bse}{\begin{subequations}}
	\newcommand{\ese}{\end{subequations}}

\graphicspath{{../}}
\usepackage{color}
\usepackage[colorlinks,bookmarks=false,citecolor=darkblue,linkcolor=red,urlcolor=blue]{hyperref}

\definecolor{darkred}{rgb}{0.7,0.0,0.0}

\definecolor{darkblue}{rgb}{0,0.02,0.45}

\definecolor{darkgreen}{rgb}{0.02,0.45,0.0}

\definecolor{violet}{rgb}{0.8,0.2,0.6}

\begin{document}
\title{Spin fluctuations, absence of magnetic order, and crystal electric field studies in the Yb$^{3+}$-based triangular lattice antiferromagnet Rb$_3$Yb(VO$_4$)$_2$}
\author{Sebin J. Sebastian }
\affiliation{School of Physics, Indian Institute of Science Education and Research Thiruvananthapuram-695551, India}
\affiliation{Ames National Laboratory, U.S. DOE, Iowa State University, Ames, IA 50011, USA}
\author{R. Kolay}
\affiliation{School of Physics, Indian Institute of Science Education and Research Thiruvananthapuram-695551, India}
\author{Abhidev. B}
\affiliation{School of Physics, Indian Institute of Science Education and Research Thiruvananthapuram-695551, India}
\author{Q.-P. Ding }
\affiliation{Ames National Laboratory, U.S. DOE, Iowa State University, Ames, IA 50011, USA}
\author{Y. Furukawa}
\affiliation{Ames National Laboratory, U.S. DOE, Iowa State University, Ames, IA 50011, USA}
\affiliation{Department of Physics and Astronomy, Iowa State University, Ames, IA 50011, USA}
\author{R. Nath}
\email{rnath@iisertvm.ac.in}
\affiliation{School of Physics, Indian Institute of Science Education and Research Thiruvananthapuram-695551, India}
\date{\today}
	
\begin{abstract}
We report a comprehensive experimental investigation of the structural, thermodynamic, static, and dynamic properties of a triangular lattice antiferromagnet Rb$_3$Yb(VO$_4$)$_2$. Through the analysis of magnetic susceptibility, magnetization, and specific heat, complemented by crystal electric field (CEF) calculations, we confirm the Kramers' doublet with effective spin $J_{\rm{eff}}=1/2$ ground state. Magnetic susceptibility and isothermal magnetization analysis reveal a weak antiferromagnetic interaction among the $J_{\rm{eff}}=1/2$ spins, characterized by a small Curie-Weiss temperature ($\theta_{\text{CW}}^{\text{LT}}\simeq-0.26$~K) or a reduced exchange coupling ($J/k_{\rm B} \simeq 0.18$~K).
The $^{51}$V NMR spectra and spin-lattice relaxation rate ($1/T_1$) show no evidence of magnetic long-range-order down to 1.6~K but reflects strong influence of CEF excitations in the intermediate temperatures.
At low temperatures, $1/T_1(T)$ shows pronounced frequency dependence and $1/T_1$ vs field in different temperatures follows the scaling behaviour, highlighting the role of paramagnetic fluctuations.
The CEF calculations using the point charge approximation divulge a large energy gap ($\sim 18.61$~meV) between the lowest and second lowest energy doublets, further establishing Kramers' doublet as the ground state. Our calculations also reproduce the experimental magnetization and specific heat data and indicate an in-plane magnetic anisotropy. These findings position Rb$_3$Yb(VO$_4$)$_2$ as an ideal and disorder-free candidate to explore intrinsic quantum fluctuations and possible quantum spin-liquid physics in a Yb$^{3+}$-based triangular lattice antiferromagnet.
\end{abstract}

\maketitle

\section{Introduction}
Quantum spin liquid (QSL) represents a novel quantum state of matter characterized by fractionalized excitations, long-range quantum entanglement, and absence of magnetic long-range order (LRO) even at absolute zero temperature~\cite{Balents464,*Savary016502}. Such an exotic phenomenon commonly emerge in frustrated magnets, particularly those featuring low-spins. Among these, the spin-$1/2$ geometrically frustrated two-dimensional (2D) triangular lattice antiferromagnets (TLAFs) have been identified as a cradle for exploring the interplay of magnetic frustration, quantum fluctuations, and potential QSL physics. As initially proposed by Anderson, spin-$1/2$ TLAFs were anticipated to exhibit resonating valence bond states, serving as a prototypical example of QSL~\cite{Anderson153,Bhattacharya060403}. Subsequent theoretical studies revealed that, contrary to initial expectations, an isotropic spin-$1/2$ TLAF may stabilize a non-collinear $120^{\circ}$ N\'eel state~\cite{Chubukov69,Capriotti3899}. Nonetheless, perturbations such as magnetic anisotropy, site disorder, and next nearest neighbor interactions significantly enrich the phase diagram. These effects can stabilize unconventional quantum states, including QSL~\cite{Li107202,Kundu117206,Iqbal144411,Zhu041105}, spin-nematic phases~\cite{Seifert195147}, intertwined dipolar-multipolar orders etc~\cite{Shen4530}.

Lately, Yb$^{3+}$ ($4f^{13}$)-based TLAFs have garnered substantial attention due to their intricate interplay of strong spin-orbit coupling (SOC) and crystal electric field (CEF) effects. In such systems, the total angular momentum $J=7/2$ manifold of the Yb$^{3+}$ ion split by the CEF into four well-separated Kramers' doublets ($J_z =\pm1/2$, $\pm3/2$, $\pm5/2$, and $\pm7/2$). At sufficiently low temperatures, only the lowest-energy Kramers' doublet is thermally populated, giving rise to an effective spin-$1/2$ ($J_{\rm eff}=1/2$) ground state. The large energy gap between the ground and first excited doublets ensures the validity of this low-energy description over a broad temperature range~\cite{Li167203,Hester027201,Guchhait2025,Somesh064421,Ranjith224417}. This $J_{\rm eff} = 1/2$ state amplifies the quantum fluctuations, and when combined with the geometric frustration inherent to the TLAFs, paves the way for more exotic ground states such as QSL.
Further, the presence of strong SOC often introduces anisotropic interactions which adds another dimension, rendering the spin-lattice more complex and exciting~\cite{Thompson057203,Flynn067201,Mohanty134408}. Prominent examples include YbMgGaO$_4$ and NaYb$Ch_2$ ($Ch$ = O, S, Se), which exhibit spin-liquid-like behavior and field-induced quantum phases~\cite{Paddison117,Ranjith224417,Baenitz220409,Ranjith180401}. In several Yb$^{3+}$-based TLAF compounds, the intrinsic site mixing and disorder often complicate the interpretation of their magnetic ground states. This demands continued efforts to look for structurally pristine and disorder-free TLAFs. Moreover, rare-earth-based frustrated magnets with small exchange couplings are often considered as potential materials for achieving low temperatures via adiabatic demagnetization refrigeration technique~\cite{Treu013001,*Tokiwa42}.

The recent studies have highlighted structurally pristine rare-earth-based materials in the family $A_3$Yb(VO$_4$)$_2$ ($A$ = K, Rb, Cs), which particularly stood out due to their ideal conditions to study intrinsic quantum magnetic phenomena without disorder-related ambiguities. For instance, K$_3$(Yb/Er)(VO$_4$)$_2$ have demonstrated remarkable structural integrity without detectable disorder~\cite{Voma144411,Yahne104423}. For K$_3$Yb(VO$_4$)$_2$, structural and magnetic studies have confirmed a perfect triangular lattice of $J_{\rm eff} = 1/2$ spins with no evidence of magnetic LRO down to 0.5~K~\cite{Voma144411}. In contrast, neutron diffraction studies on K$_3$Er(VO$_4$)$_2$ ($J_{\rm eff}=1/2$) have shown the coexistence of quasi-2D and 3D magnetic orderings around $T_{\rm N}\simeq 0.15$~K~\cite{Yahne104423}, emphasizing the crucial role of single-ion anisotropy and exchange anisotropy in this system. Additionally, the structurally perfect TLAF YbZn$_2$GaO$_5$ has provided compelling evidence for Dirac QSL, further illustrating the rich spectrum of quantum phenomena accessible in these materials~\cite{Bag266703}.

\begin{figure}
\includegraphics[width=\columnwidth] {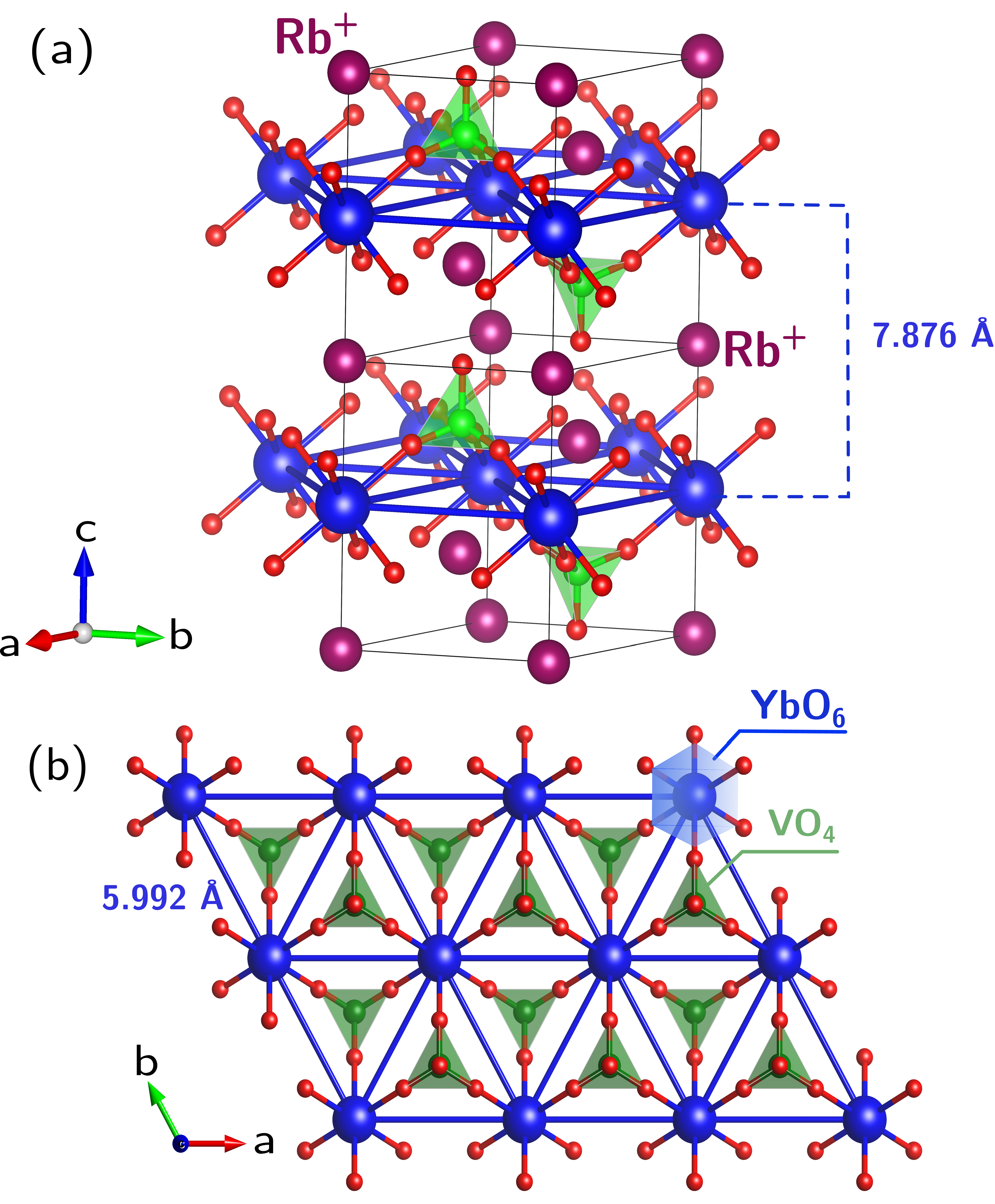}
\caption{\label{Fig1}(a) Crystal structure of Rb$_3$Yb(VO$_4$)$_2$ in a unit cell with YbO$_6$ octahedra and VO$_4$ tetrahedral units separated by Rb-atoms. (b) A section of the layer showing triangular units formed by Yb$^{3+}$, connected via VO$_4$ tetrahedra.}
\end{figure}
Herein, we present a comprehensive experimental investigation of the structural, thermodynamic, static, and dynamic properties of the TLAF Rb$_3$Yb(VO$_4$)$_2$. It crystallizes in a hexagonal structure with space group $P\bar{3}m1$. In the crystal structure, the regular YbO$_6$ octahedra are interconnected via VO$_4$ tetrahedra, forming isotropic triangular layers within the $ab$-plane [see Fig.~\ref{Fig1}(b)]. These triangular layers are separated by Rb$^{+}$ ions, as illustrated in Fig.~\ref{Fig1}(a). Our experiments complemented by CEF calculations reveal a strongly anisotropic Kramers' doublet with $J_{\rm eff} = 1/2$ ground state. The $^{51}$V NMR results provide evidence for persistent spin fluctuations without magnetic LRO down to 1.6~K. Additionally, the CEF calculations reproduce the experimental thermodynamic data, confirming significant in-plane magneto-crystalline anisotropy.

\section{Methods}
\begin{figure}
\includegraphics[width=\columnwidth]{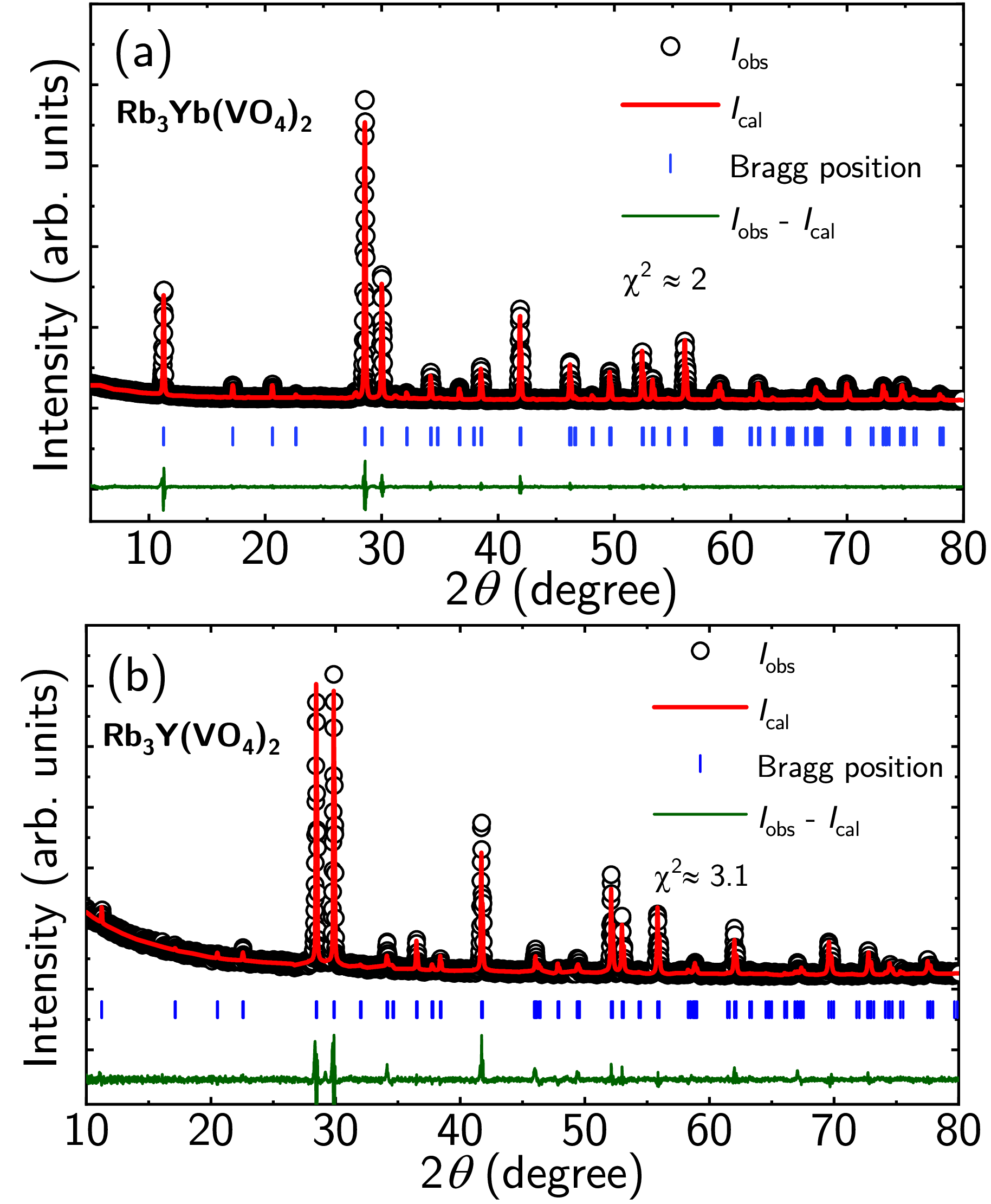}
\caption{\label{Fig2} Room temperature powder XRD pattern of (a) Rb$_3$Yb(VO$_4$)$_2$ and (b) Rb$_3$Y(VO$_4$)$_2$. The open circles represent the experimental data and the solid red line is the Rietveld refined fit. The blue vertical lines are calculated Bragg positions. The green line at the bottom indicates the difference between observed and experimental intensities.}
\end{figure}
Polycrystalline samples of Rb$_3$Yb(VO$_4$)$_2$ and its nonmagnetic analog Rb$_3$Y(VO$_4$)$_2$ were synthesized via a conventional solid-state reaction route. Stoichiometric mixtures of high-purity precursors Rb$_2$CO$_3$ (Sigma Aldrich, $\geq$ 99.9\%), Y$_2$O$_3$ (Sigma Aldrich, $\geq$ 99.9\%), Yb$_2$O$_3$ (Sigma Aldrich, $\geq$ 99.99\%), and NH$_4$VO$_3$ (Sigma Aldrich, $\geq$ 99.99\%) were thoroughly ground for several hours and pressed into pellets. These pellets were placed in a platinum crucible and subjected to a two-step heat treatment: first at 500\degree C for 12~hours, followed by a second firing at 750\degree C for 48~hours. The samples were then quenched to room temperature.

Phase purity and crystal structure were confirmed using powder x-ray diffraction (XRD) data collected in a PANalytical diffractometer employing Cu$K_\alpha$ radiation ($\lambda_{\rm avg} = 1.5418$~~\AA) at room temperature. Rietveld refinement of the powder XRD data was carried out using the \verb|FULLPROF| suite~\cite{Carvajal55}, with initial structural parameters taken from the isostructural compound K$_3$Yb(VO$_4$)$_2$~\cite{Voma144411}. The refined structure corresponds to a hexagonal unit cell with space group $P\bar{3}m1$. The refined lattice parameters are $a = b = 5.951(2)$~\AA, $c = 7.853(1)$~\AA, $\beta = 120(1)^\circ$, and unit cell volume $V_{\rm cell} \simeq 241.61$~\AA$^3$ for Rb$_3$Yb(VO$_4$)$_2$ and $a = b = 5.987(1)$~\AA, $c = 7.888(1)$~\AA, $\beta = 120(1)^\circ$, and $V_{\rm cell} \simeq 244.86$~\AA$^3$ for Rb$_3$Y(VO$_4$)$_2$. Similarly, the refined atomic positions are summarized in Table~\ref{Refined parameters}. No extra peaks associated with any foreign phases was detected in the powder XRD.
\begin{table}[ptb]
\caption{The Wyckoff positions and refined atomic coordinates for each atom of Rb$_3$(Yb/Y)(VO$_4$)$_2$ at room temperature, obtained from the powder XRD refinement.}
\label{Refined parameters}
\begin{ruledtabular}
\begin{tabular}{ccccccc}	
\multicolumn{1}{p{1cm}}{\centering Atomic \\ sites} & \multicolumn{1}{p{1.3 cm}}{\centering Wyckoff\\ positions} &\multicolumn{1}{p{1.5cm}}{\centering $x$} &\multicolumn{1}{p{1.3cm}}{\centering $y$} &\multicolumn{1}{p{1.3 cm}}{\centering $z$}  &  Occ.
\\\hline
Yb(1) &  $1b$ & 1 & 0 & 0.5 & 1 \\
Y(1) &       & 1 & 0 & 0.5 & 1 \\
Rb(1) & $2d$ & 0.6667(1) & 0.3333(1) & 1.2877(1)  & 1 \\
      &     & 0.6667(1) & 0.3333(1) & 1.2754(2)  & 1 \\
Rb(2) & $1a$ & 0 & 0 & 1  & 1 \\
     &    & 0 & 0 & 1  & 1 \\
V(1) & $2d$ & 0.6667(1) & 0.3333(1) & 0.7374(1) & 1 \\
     &   & 0.6667(1) & 0.3333(1) & 0.7702(2) & 1 \\
O(1) & $6i$ & 0.3867(1) &  0.1934(1) & 0.6658(1) &1\\
     &   & 0.3870(1) &  0.1935(1) & 0.6483(2) &1\\
O(2) & $2d$ & 0.6667(1) &  0.3333(1) & 0.9470(1) &1\\
     &   & 0.6667(1) &  0.3333(1) & 0.9755(2) &1
\\ \hline
\end{tabular}
\end{ruledtabular}
\end{table}

Magnetization ($M$) measurements as a function of temperature ($T$) were performed in the range 0.4~K~$\leq T \leq$~380~K under applied magnetic fields ($H$) using a superconducting quantum interference device (SQUID) magnetometer (MPMS-3, Quantum Design). For measurements below 1.8~K, a $^3$He insert (\verb|iHelium3|) was employed. Isothermal magnetization ($M$ vs $H$) data were collected at different temperatures in magnetic fields up to 7~T. Temperature-dependent specific heat, $C_{\rm P}(T)$ was measured on a sintered pellet in the temperature range 1.9~K~$\leq T \leq$~300~K and in magnetic fields up to 9~T using the standard thermal relaxation method in a Physical Property Measurement System (PPMS, Quantum Design).

Nuclear magnetic resonance (NMR) experiments were carried out using a laboratory-built, phase-coherent spin-echo spectrometer on the $^{51}$V nucleus ($I = 7/2$) with a gyromagnetic ratio $\gamma_{\rm N}/2\pi = 11.19$~MHz/T over the temperature range $1.6-300$~K. Spectra were recorded at fixed radio frequencies by sweeping the magnetic field, utilizing a standard $\pi/2 - \tau - \pi$ pulse sequence with $\tau = 20~\mu$s. The nuclear spin-lattice relaxation rate ($1/T_1$) was determined by monitoring the recovery of longitudinal magnetization using the saturation recovery method ($\pi/2-\tau_1 - \pi/2 - \tau_2 - \pi$ sequence), at the central peak position. 

Crystal electric field (CEF) calculations for Rb$_3$Yb(VO$_4$)$_2$ were performed within the point charge approximation using the \verb|PyCrystalField| Python package~\cite{Scheie356}. The required structural parameters, including lattice constants and atomic positions, were taken from the powder XRD refinement (Table~\ref{Refined parameters}). The calculation explicitly included the six nearest-neighbor O$^{2-}$ ions forming the YbO$_6$ octahedron, ensuring an accurate estimation of the local crystal field environment at the Yb$^{3+}$ site.

\section {Results}
\subsection{Magnetization}
\begin{figure}
\includegraphics[width=\columnwidth]{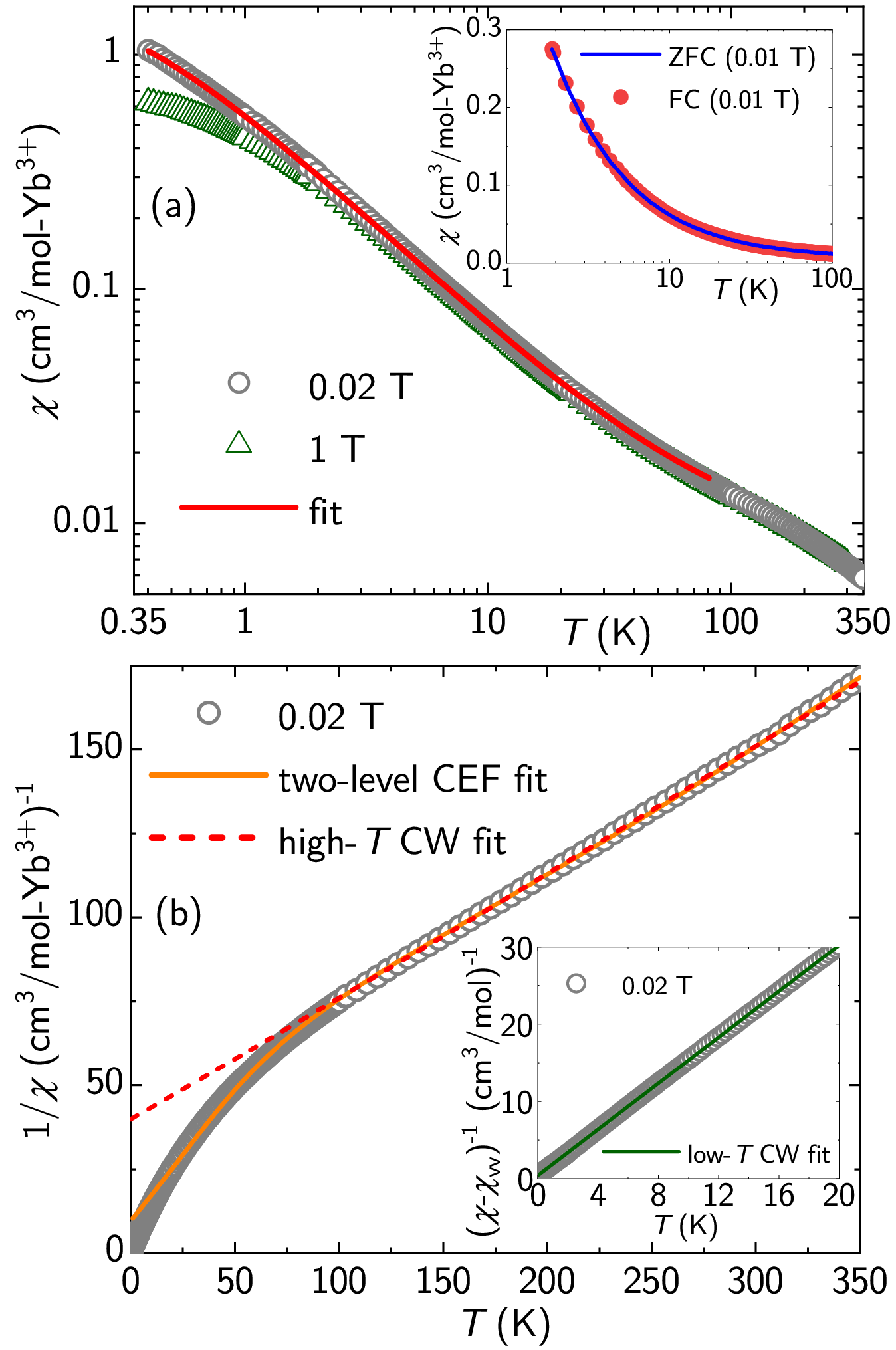}
\caption{\label{Fig3} (a) $\chi(T)$ measured in applied magnetic fields of $\mu_{0}H$ = 0.02~T and 1~T. The solid line represents the fit mentioned in the text. Inset: $\chi(T)$ measured at $\mu_{0}H$ = 0.01 T in both ZFC and FC protocols. (b) Inverse susceptibility ($1/\chi$) vs $T$ measured at $\mu_0H = 0.02$~T. The solid and dashed lines represent the high-$T$ CW fit [Eq.~\eqref{CW}] for $T \geq 150$~K and two-level CEF fit [Eq.~\eqref{CEF_CW}] for $T \geq 25$~K, respectively. Both fits are extrapolated down to low temperatures. Inset: The low-$T$ $1/(\chi-\chi_{\rm VV}$) data along with the CW fit.}
\end{figure}
Temperature dependent magnetic susceptibility, $\chi[\equiv M/H$], of Rb$_3$Yb(VO$_4$)$_2$ measured in various magnetic fields is shown in Fig.~\ref{Fig3}(a). $\chi(T)$ follows Curie-Weiss (CW) behavior at high temperatures and no magnetic LRO is evidenced down to 0.4~K. The inset of Fig.~\ref{Fig3}(a) displays $\chi(T)$ measured in 100~Oe under both zero-field-cooled (ZFC) and field-cooled (FC) protocols, showing no bifurcation down to 1.8~K. This absence of ZFC-FC separation excludes the possibility of spin-glass behavior or spin freezing at low temperatures. Figure~\ref{Fig3}(b) presents the inverse susceptibility $1/\chi$ as a function of temperature. A clear slope change below $\sim 100$~K reflects the depopulation of excited CEF levels. For $T > 150$~K, $\chi(T)$ was fitted well by a modified CW law
\begin{equation}
\chi(T) = \chi_0 + \frac{C}{T - \theta_{\rm CW}},
\label{CW}
\end{equation}
where, $\chi_0$ accounts for $T$-independent core diamagnetic and Van-Vleck paramagnetic contributions, $C$ is the Curie constant, and $\theta_{\rm CW}$ is the CW temperature. The fit yields $\chi_0^{\rm HT} \simeq -2.56(1) \times 10^{-4}$~cm$^3$/mol, $C^{\rm HT} \simeq 2.82(1)$~cm$^3$K/mol, and $\theta_{\rm CW}^{\rm HT} \simeq -109.7(2)$~K. From the value of $C^{\rm HT}$, the effective magnetic moment is estimated as $\mu_{\rm eff}^{\rm HT} = \sqrt{8C}\mu_{\rm B} \simeq 4.74(1)~\mu_{\rm B}$, which is in good agreement with the expected value $\mu_{\rm eff} \simeq 4.54~\mu_{\rm B}$ for a free Yb$^{3+}$ ion ($J = 7/2$, $g_J = 1.14$) in a $4f^{13}$ configuration. The large and negative $\theta_{\rm CW}^{\rm HT}$ does not necessarily indicate strong AFM interactions, rather reflects the CEF effects due to thermal occupation of excited states.
Note that at high temperatures, the high energy doublets contribute significantly to susceptibility with a sizable Van-Vleck contribution~\cite{Somesh064421,Ranjith224417}.

To understand the low-$T$ magnetic behavior, $1/\chi(T)$ was replotted after subtracting the Van-Vleck contribution [i.e., $1/(\chi - \chi_{\rm VV})$], as shown in the inset of Fig.~\ref{Fig3}(b). The value of $\chi_{\rm VV}$ was estimated from the magnetization isotherm at $T = 0.4$~K (discussed later). A CW fit in the temperature range 2–20~K yields $C^{\rm LT} \simeq 0.67(2)$~cm$^3$K/mol and $\theta_{\rm CW}^{\rm LT} \simeq -0.27(1)$~K. This value of $C^{\rm LT}$ corresponds to an effective moment $\mu_{\rm eff}^{\rm LT} \simeq 2.31(1)~\mu_{\rm B}$, suggestive of an effective spin $J_{\rm eff} = 1/2$ with an average $g$-value of $g_{\rm ave} \simeq 2.67(1)$, typically expected for Yb$^{3+}$ based systems~\cite{Voma144411}. A small value of $\theta_{\rm CW}^{\rm LT}$ indicates very weak AFM interactions among the $J_{\rm eff} = 1/2$ moments. Using the mean-field approximation for a TLAF with coordination number $z = 6$, the exchange coupling is estimated to be $J/k_{\rm B} = 2\theta_{\rm CW}^{\rm LT}/3 \simeq 0.18(1)$~K~\cite{Sebastian104425}.

Since the Yb$^{3+}$ ions form an isotropic triangular lattice and the CEF stabilizes an effective $J_{\rm eff}=1/2$ Kramers' doublet ground state, we employed the high-temperature series expansion for an $S = 1/2$ isotropic triangular lattice antiferromagnet to fit the $\chi(T)$ data at low temperatures. This uses the Pad\'e approximation and has the form~\cite{Masafumi47}
\begin{equation}
\chi(T) = \frac{N_{\rm A} \mu_0 g^2 \mu_{\mathrm{B}}^2}{4 k_{\mathrm{B}} T}
\left.\frac{1 + b_1 x + \dots + b_6 x^6}{1 + c_1 x + \dots + c_7 x^7}\right|_{x=J/(4k_{\mathrm{B}}T)}.
\label{HTSE}
\end{equation}
Here, $N_{\rm A}$ is the Avogadro's number, $k_{\rm B}$ is the Boltzmann constant, and the Pad\'e coefficients $b_i$ and $c_i$ are taken from Ref.~\cite{Elstner1629}. As shown in Fig.~\ref{Fig3}(a), below 100~K, $\chi(T)$ is well captured by Eq.~\eqref{HTSE}, yielding $J/k_{\rm B} \simeq 0.19(2)$~K and $g \simeq 2.67(1)$. The departure of this model from the experimental $\chi(T)$ data above 100~K is attributed to the thermal population of higher CEF levels~\cite{Guchhait144434}. This value of $J/k_{\rm B}$ appears to be in good agreement with the value estimated from $\theta_{\rm CW}^{\rm LT}$.

To estimate the energy scale of the CEF excitations, the high-$T$ $\chi(T)$ data were also modeled using an effective two-level scheme~\cite{Mugiraneza95}
\begin{equation}
\chi(T) = \chi_0 + \frac{1}{8\left( T-\theta_{\rm CW}^{\rm CEF} \right)} 
\left[\frac{ \mu_{\rm eff,0}^2 + \mu_{\rm eff,1}^2 \, e^{ -\Delta^{\rm CEF} / k_{\rm B} T } }{ 1 + e^{ -\Delta^{\rm CEF} / k_{\rm B} T } }
\right],
\label{CEF_CW}
\end{equation}
where $\mu_{\rm eff,0}$ and $\mu_{\rm eff,1}$ are the effective moments of the ground and first excited doublets, respectively, and $\Delta^{\rm CEF}$ is the energy gap between them. This model fit for $T \geq 25$~K yields $\chi_0^{\rm CEF} \simeq -1.62(1) \times 10^{-3}$~cm$^3$/mol, $\mu_{\rm eff,0}^{\rm CEF} \simeq 3.28(1)~\mu_{\rm B}$, $\mu_{\rm eff,1}^{\rm CEF} \simeq 4.26(1)~\mu_{\rm B}$, $\Delta^{\rm CEF}/k_{\rm B} \simeq 201(3)$~K, and $\theta_{\rm CW}^{\rm CEF} \simeq -12.60(2)$~K. Here, the value of $\Delta^{\rm CEF}/k_{\rm B}$ is comparable with the one obtained from the point charge calculations (discussed later). However, the value of $\theta_{\rm CW}^{\rm CEF}$ is significantly different from $\theta_{\rm CW}^{\rm LT}$. When the fit is done covering the entire temperature range (0.4~K$\leq T \leq 380$~K), the obtained parameters [$\theta_{\rm CW}^{\rm CEF} \simeq -1.3(1)$~K and $\Delta^{\rm CEF}/k_{\rm B} \simeq 125(2)$~K] are drastically different from $\theta_{\rm CW}^{\rm LT}$ and $\Delta^{\rm CEF}/k_{\rm B}$ from the point charge calculations~\cite{Arjun014013,Arjun224415,Pula014412}.
This discrepancy can be attributed to the fact that, this model neglects the Van-Vleck contribution and assumes purely Curie-type behavior for each doublet. This makes the model inadequate to capture the full CEF response in $\chi(T)$.

\begin{figure}
\includegraphics[width=\columnwidth]{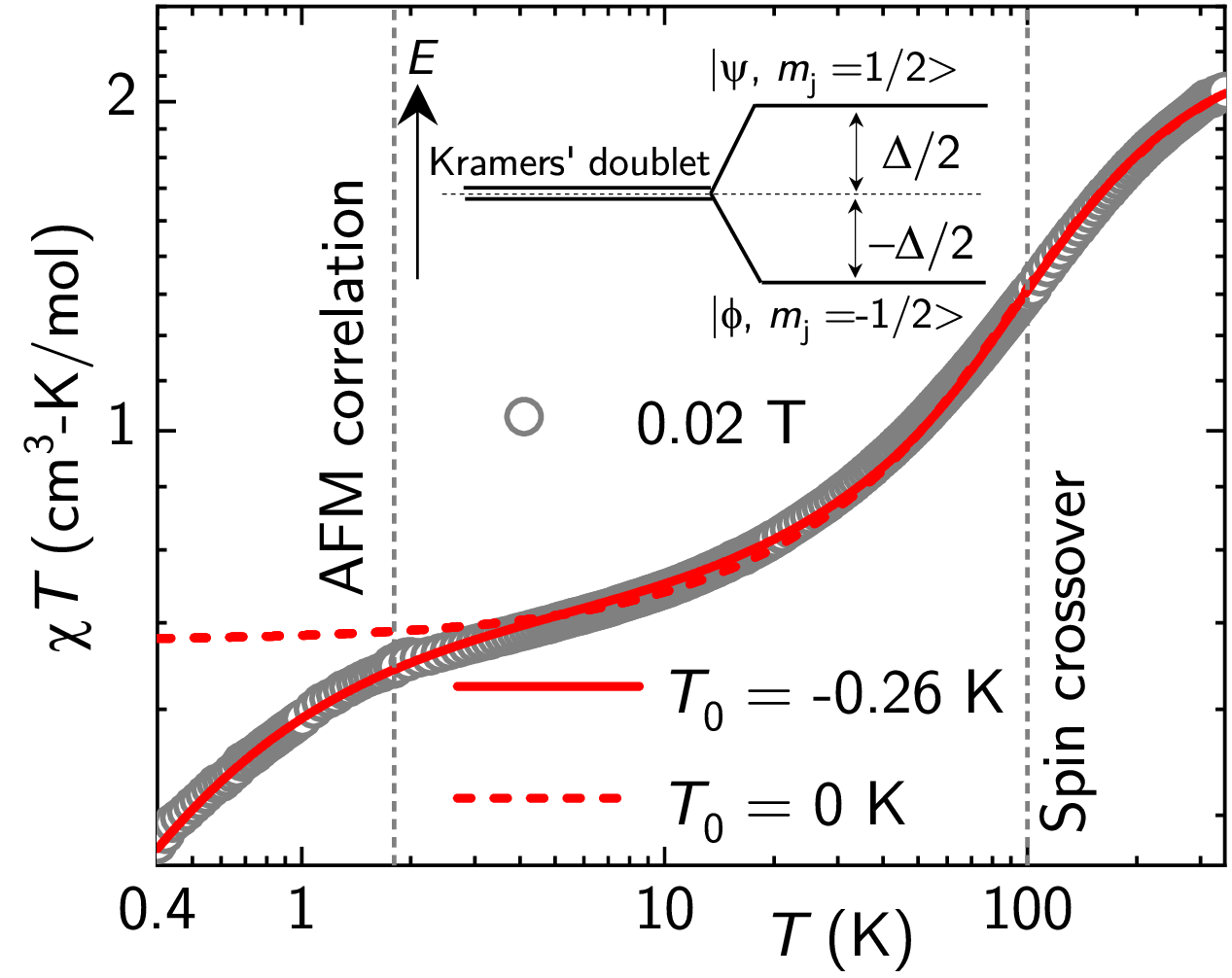}
\caption{\label{Fig4} $\chi T$ vs $T$ for $\mu_0H = 0.02$~T. The solid and dashed lines are fits using Eq.~\eqref{VV_KD_new} for $T_0 = 0$ and $T_0 = -0.26$~K, respectively. Inset: Schematic energy diagram of the Zeeman split levels of the lowest Kramers' doublet.}
\end{figure}
In order to have an accurate description of $\chi(T)$ over the entire temperature range, we used the following modified two-level scheme that accounts for both Curie and Van-Vleck terms (see Appendix ~\ref{appx})~\cite{Sebastian034403}
\begin{equation}
\begin{split}
\chi = \chi_0 +\frac{N}{k_B (T-T_0)} \Bigg[ 
		& C_0 
		+  C_1\left( \frac{2k_B T}{\Delta} \right) 
		\tanh\left( \frac{\Delta}{2k_B T} \right) \\
		& + C_2 \tanh\left( \frac{\Delta}{2k_B T} \right) 
		\Bigg].
	\end{split}
    \label{VV_KD_new}
\end{equation}
Here, the introduced $T_0$ is analogous to $\theta_{\rm CW}$. As shown in Fig~\ref{Fig4}, we are able to reproduce the entire temperature profile of $\chi T$ with a finite value of $T_0$. When $T_0$ is fixed to zero, it fits the data well in the high-$T$ range and stays constant below about 10~K, as expected. The fit resulted in $\chi_0 \simeq -1.41(1) \times 10^{-4}$~cm$^3$/mol, $T_0 \simeq -0.26(1)$~K, $C_0/k_{\rm B} \simeq 1.79(2)$~cm$^3$-K/mol, $C_1/k_{\rm B} \simeq 0.931(1)$~cm$^3$-K/mol, $C_2/k_{\rm B} \simeq -1.044(2)$~cm$^3$-K/mol, and $\Delta/k_{\rm B}\simeq 262.3(2)$~K. This value of $T_0$ agrees very well with $\theta_{\rm CW}^{\rm LT}$. In the low-$T$ regime where $T\ll\Delta_{\rm CEF}/k_{\rm B}$, the second term within the parentheses in Eq.~\eqref{VV_KD_new} is almost negligible. The terms $C_0/k_{\rm B}$ and $C_2/k_{\rm B}$ share similarity with the Curie constant from Eq.~\eqref{CW}. We observed that the effective Curie constant, $C_{\rm eff} = C_0+C_2 \simeq 0.755$~cm$^3$-K/mol corresponds to $\mu_{\rm eff} \simeq 2.45~\mu_{\rm B}$, which appears very close to $\mu_{\rm eff}^{\rm LT}$ obtained from the low-$T$ $\chi(T)$ analysis.

\begin{figure}
\includegraphics[width=\columnwidth]{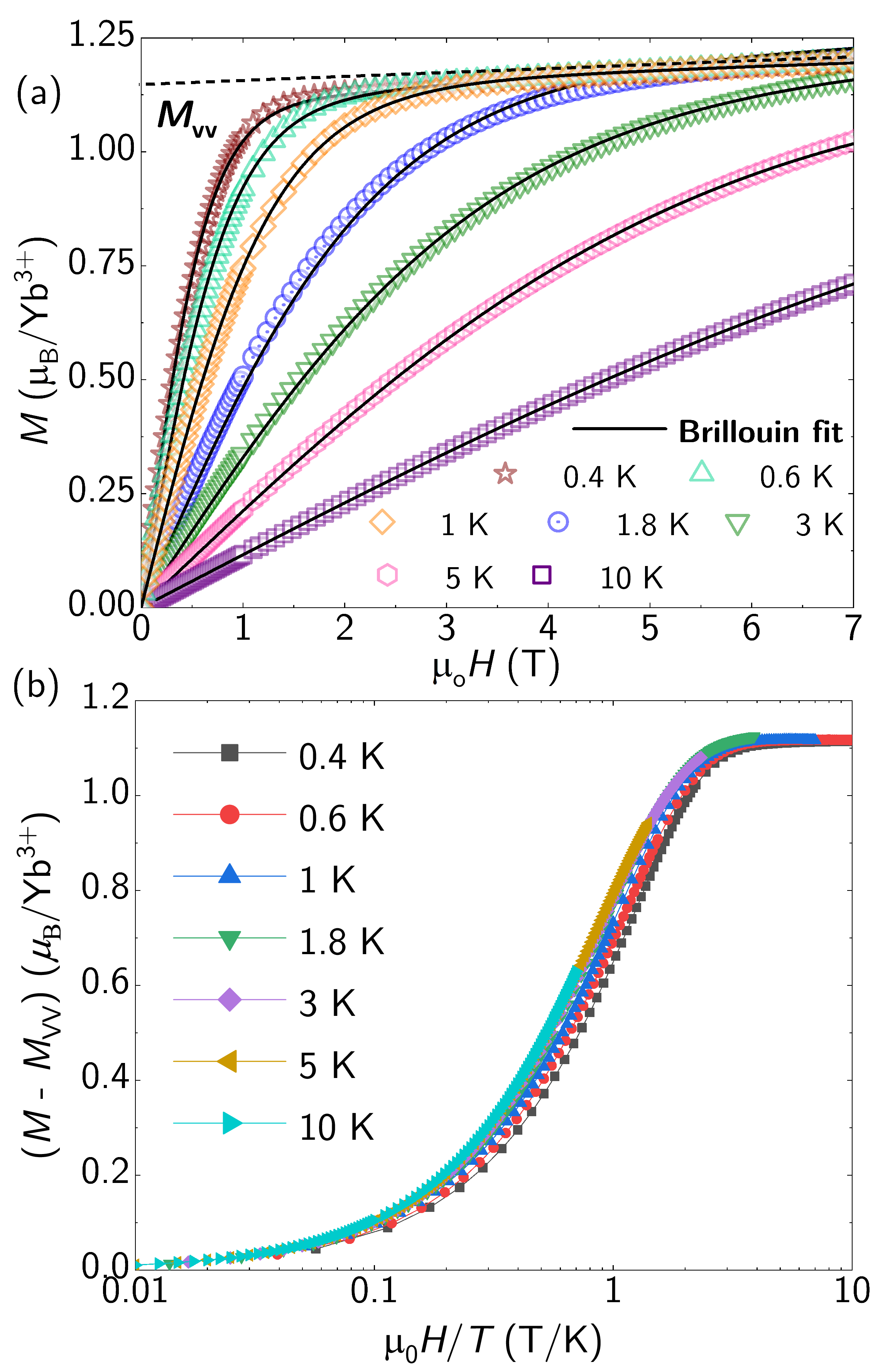}
\caption{\label{Fig5} (a) $M$ vs $H$ at different temperatures along with the Brillouin fits. The dashed line represents a linear fit to the high-field data at $T = 0.4$~K. (b) ($M-M_{\rm VV}$) vs $\mu_{0}H/T$ at different temperatures. The $x$-axis is shown in log scale to show the deviation below 1~K more clearly.}
\end{figure}
Figure~\ref{Fig5}(a) presents the isothermal magnetization $M(H)$ of Rb$_3$Yb(VO$_4$)$_2$ measured at various temperatures. At $T = 0.4$~K, the $M(H)$ curve exhibits saturation near $\mu_0H_{\rm S} \simeq 1.1$~T, followed by a slow and linear increase at higher fields. This linear increase reflects the characteristics of the Van-Vleck paramagnetic contribution arising from the
field-induced electronic transitions~\cite{Li167203}. To quantify this, we did a linear fit to the data in the high-field region ($\mu_0H > 5$~T). From the slope and $y$-intercept of this fit, we estimate the Van-Vleck susceptibility $\chi_{\rm VV} \simeq 6.6(1) \times 10^{-3}$~cm$^3$/mol and the saturation magnetization $M_{\rm S} \simeq 1.15(1)~\mu_{\rm B}$ per Yb$^{3+}$ ion, respectively. This saturation value corresponds to $J_{\rm eff}=1/2$ state with an average $g$-value of $g_{\rm avg} \simeq 2.3$, close to the value inferred from the $\chi(T)$ analysis.

As discussed earlier, the low-$T$ CW fit yields a small and negative $\theta_{\rm CW}^{\rm LT}$, suggesting weak AFM interactions. In light of this, we modeled the magnetization isotherms assuming non-interacting paramagnetic moments using the following expression~\cite{Sebastian034403}
\begin{equation}
\label{MH imp}
M(H) = \chi_{\rm VV} H + N_{\rm A} g \mu_{\rm B} J_{\rm eff} B_{J_{\rm eff} }(x),
\end{equation}
where $B_{J_{\rm eff} }(x)$ is the Brillouin function and $x = g \mu_{\rm B} J_{\rm eff} H / (k_{\rm B} T)$. For $J_{\rm eff} = 1/2$, the Brillouin function simplifies to $B_{J_{\rm eff} }(x) = \tanh(x)$~\cite{Kittel1986}. During fitting, we fixed $J_{\rm eff} = 1/2$ and $\chi_{\rm VV}$ to the values obtained earlier, leaving $g$ as the only free parameter. The experimental isotherms at low temperatures are well described by Eq.~\eqref{MH imp}, supporting the scenario of weakly interacting paramagnetic $J_{\rm eff} = 1/2$ moments. The fits yield $g_{\rm avg} \simeq 2.3(1)$, in agreement with the values obtained from both the saturation magnetization and low-$T$ $\chi(T)$ analysis.

To further confirm the paramagnetic behavior, we subtracted the Van-Vleck contribution from $M(H)$ and plotted $M - M_{\rm VV}$ as a function of $\mu_0H/T$ in Fig.~\ref{Fig5}(b)~\cite{Somesh064421}. Above 1~K, all the curves collapse onto a universal curve, consistent with paramagnetic scaling expected for non-interacting spins. However, a noticeable deviation from this scaling appears below 1~K, suggesting the onset of finite AFM correlations in this temperature regime.

\subsection{Specific heat}
\begin{figure*}
\includegraphics[width=\textwidth, height = 5.5cm]{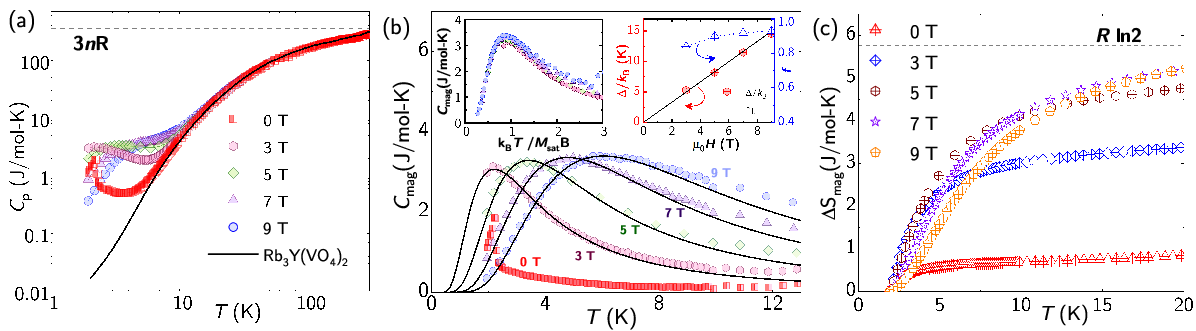}
\caption{\label{Fig6} Temperature dependence of $C_{\rm p}$ for Rb$_3$Yb(VO$_4$)$_2$ measured in various applied magnetic fields. The solid line denotes $C_{\rm ph}(T)$ of non-magnetic analogue Rb$_3$Y(VO$_4$)$_2$, while the horizontal dashed line marks the Dulong–Petit limit. (b) $C_{\rm mag}(T)$ obtained after phonon subtraction, shown for different magnetic fields along with fits by a two-level Schottky model. Insets: (left) Scaled plot of $C_{\rm mag}$ vs $k_{\rm B}T/(M_{\rm sat}B)$ demonstrating the data collapse at different fields. (right) Field dependence of the energy gap $\Delta/k_{\rm B}$ (left $y$-axis) and $f$ (right $y$-axis). The solid line represents a linear fit to $\Delta/k_{\rm B}(H)$. (c) $S_{\rm mag}(T)$ for different applied fields.}
\end{figure*}
Temperature-dependent specific heat [$C_{\rm p}(T)$] of Rb$_3$Yb(VO$_4$)$_2$ measured down to 2~K at various applied magnetic fields is presented in Fig.~\ref{Fig6}(a). Similar to $\chi(T)$, no indication of magnetic LRO is observed down to 2~K. A pronounced $\lambda$-type anomaly at $\sim 2.23$~K in the zero-field data corresponds to the magnetic ordering from a trace amount of Yb$_2$O$_3$ impurity. It is to be noted that a trace amount of unreacted Yb$_2$O$_3$ is typically found in most of the Yb$^{3+}$-based compounds~\cite{Arjun014013,Arjun224415}. Since the specific heat anomaly of Yb$_2$O$_3$ is particle size dependent, in some of the compound, it goes unnoticed~\cite{Rojas879}. In this case, as the extrinsic phase is untraceable in powder XRD, we presume that its phase fraction would be about 1~\%. As the magnetic field is applied, a broad maximum appears that shifts towards high temperatures with field, a fingerprint of Schottky anomaly appearing due to the virtual transitions among the Zeeman split levels.

In magnetic insulators, $C_{\rm p}(T)$ is composed of two distinct contributions: the phonon or lattice contribution [$C_{\rm ph}(T)$], which dominates at higher temperatures and the magnetic specific heat [$C_{\rm mag}(T)$], prevalent at lower temperatures. To isolate $C_{\rm mag}(T)$, we measured specific heat of the non-magnetic analog, Rb$_3$Y(VO$_4$)$_2$ which serves as $C_{\rm ph}$. After mass correction, $C_{\rm ph}(T)$ matches well with the high-temperature $C_{\rm p}(T)$ data of Rb$_3$Yb(VO$_4$)$_2$ [see the solid line in Fig.~\ref{Fig6}(a)]. Upon subtracting this lattice contribution, we successfully extracted $C_{\rm mag}(T)$. The resulting $C_{\rm mag}(T)$ in different magnetic fields is displayed in Fig.~\ref{Fig6}(b). The magnetic entropy [$S_{\rm mag}(T)$] calculated by integrating $C_{\rm mag}/T$ with respect to temperature is shown in Fig.~\ref{Fig6}(c). In zero-field, we could only recover a small fraction of the total entropy expected for a two-level Kramers' doublet system ($R\ln 2 = 5.76$~Jmol$^{-1}$K$^{-1}$) since a large fraction of entropy is released in low temperatures ($T < 2$~K) due to the development of magnetic correlations. In higher fields, as the magnetic correlation is suppressed and Schottky anomaly shifts to high temperatures, the accumulation of entropy also moves towards higher temperatures. This enables us to recover the full entropy at higher fields. Indeed, for $\mu_0 H > 5$~T, we achieved nearly $S_{\rm mag}(T) \simeq 5.25$~Jmol$^{-1}$K$^{-1}$ at around 20~K which accounts roughly 91.1\% of the expected entropy ($R\ln 2 = 5.76$~Jmol$^{-1}$K$^{-1}$). This further supports a $J_{\rm eff} = 1/2$ ground state at low temperatures for Rb$_3$Yb(VO$_4$)$_2$.


To quantitatively analyze this Schottky anomaly, $C_{\rm mag}(T)$ was fitted using a two-level Schottky model
\begin{equation}\label{Schottky}
C_{\rm Sch} (T, H) = fR\left(\frac{\Delta}{k_{\rm B}T}\right)^2\frac{e^{\left(\frac{\Delta}{k_{\rm B}T}\right)}}{\left[e^{\left(\frac{\Delta}{k_{\rm B}T}\right)}+1\right]^{2}}.
\end{equation}
Here, $f$ represents the molar fraction of free spins and $\Delta/k_{\rm B}$ denotes the energy gap between the Zeeman-split ground state levels. Excellent agreement between the experimental data and the Schottky fits [Fig.~\ref{Fig6}(b)] enabled extraction of the parameters $f$ and $\Delta/k_{\rm B}$, displayed in the right inset of Fig.~\ref{Fig6}(b). The parameter $f$ progressively increases with the applied field, approaching saturation ($f\approx 1$) at higher fields, signifying full excitation of Yb$^{3+}$ spins. Conversely, at fields below saturation, a fraction of spins remain correlated that diminishes with increasing field strength. The extracted energy gap, $\Delta/k_{\rm B}$, varies linearly with applied field, yielding a small zero-field gap of $\Delta/k_{\rm B}(0)\approx 1.66(1)$~K. Furthermore, using the measured $\Delta/k_{\rm B}\approx 14.5$~K at 9~T, the Land\'e $g$-factor ($g=\Delta/\mu_{\rm B}H$) is estimated to be about 2.4, consistent with the magnetization analysis. 

A scaling analysis of $C_{\rm mag}$ vs $k_{\rm B}T/(M_{\rm sat}\mu_0H)$ is presented in the left inset of Fig.~\ref{Fig6}(b). Remarkably, the data for fields $\mu_0 H \geq 2$~T collapse onto a universal curve, highlighting weak magnetic interactions among the Yb$^{3+}$ ions and also corroborate the reduced value of $\theta_{\rm CW}^{\rm LT}$~\cite{Guo094404,Guchhait2025}.

\subsection{Nuclear Magnetic Resonance}
\begin{figure}
\includegraphics[width=\linewidth]{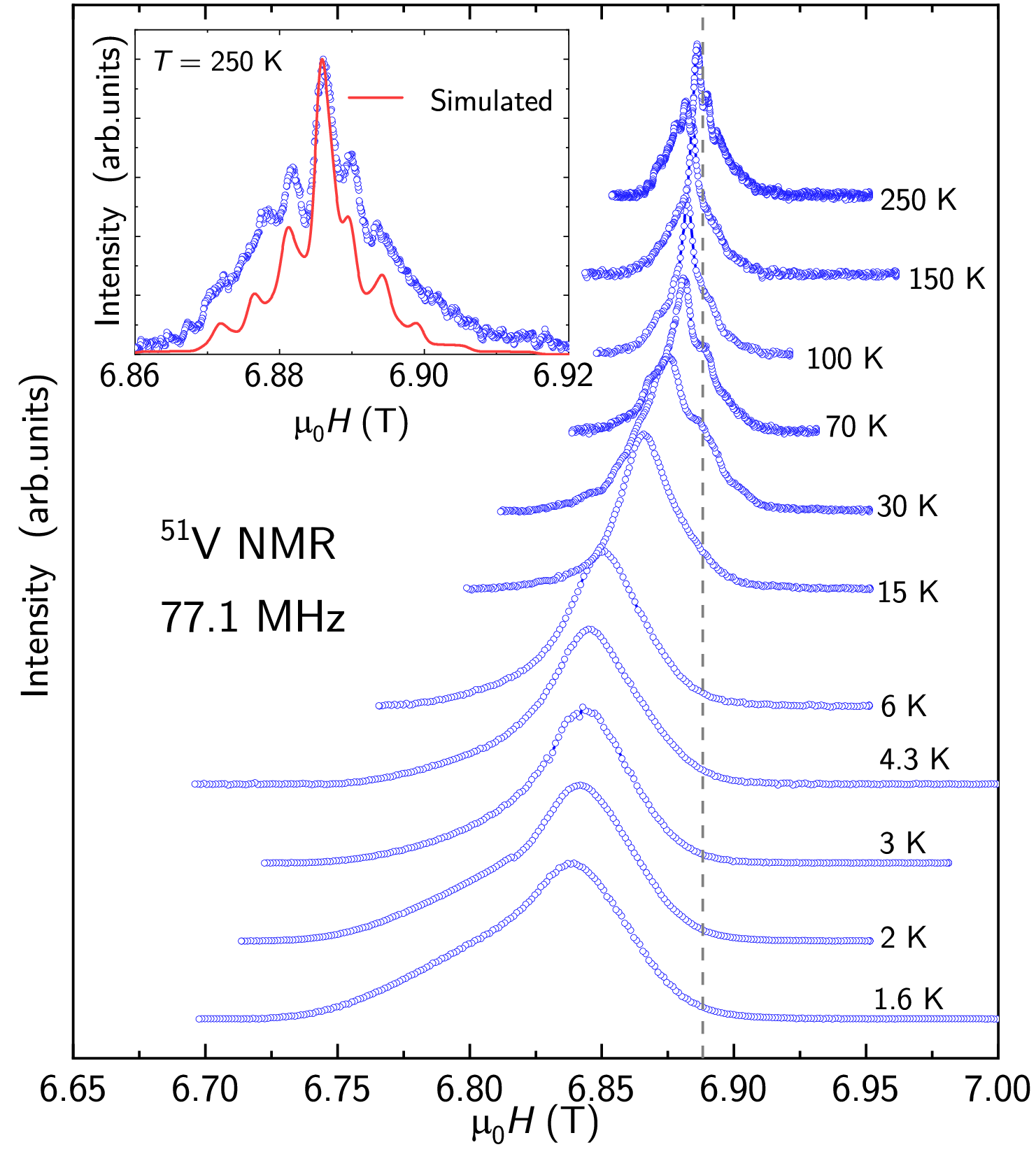}
\caption{\label{Fig7} Typical field-sweep $^{51}$V NMR spectra of Rb$_3$Yb(VO$_4$)$_2$ measured at 77.1~MHz at various temperatures. The dashed line denotes the Larmor field. Inset: $^{51}$V spectrum at $T = 250$~K and the solid red line is the simulated spectrum.}
\end{figure}
NMR is a powerful local technique used to investigate both the static and dynamic properties of spin systems. In Rb$_3$Yb(VO$_4$)$_2$, the $^{51}$V nuclei are coupled hyperfinely to the magnetic Yb$^{3+}$ ions arranged in a triangular geometry. This strong hyperfine coupling enables detailed probing of the low-energy spin excitations of the Yb$^{3+}$ ions through $^{51}$V NMR measurements. Further, the $^{51}$V nuclei ($I = 7/2$) being in a non-cubic local symmetry, one expects quadrupole interactions between the $^{51}$V moment and electric field gradient (EFG) at the $^{51}$V position. This quadrupolar interaction may partially lifts the degeneracy of the nuclear spin states, thereby influencing the nuclear resonance frequency. The complete nuclear spin Hamiltonian, comprising both the Zeeman interaction (due to an applied external magnetic field) and the quadrupolar interaction can be expressed as~\cite{Curro026502,Slichter1992}
\begin{equation}
\mathcal{H} = -\gamma \hbar \hat{I}H(1+K) + \frac{h\nu_Q}{6}\left[(3\hat{I}_{z}^2 - \hat{I}^2)+\eta(\hat{I}_{x}^2 - \hat{I}_{y}^2)\right],
\label{Eq8}
\end{equation}
where $\nu_{\rm Q}= \frac{3e^2qQ}{2I(2I-1)h}$ is the nuclear quadrupole resonance frequency. Here, $e$ is the electron charge, $\hbar = h/2\pi$ the reduced Planck's constant, $H$ is the external magnetic field along $\hat{z}$, $K$ is the magnetic shift due to the hyperfine field at the nuclear site, and $\eta$ describes the asymmetry of the EFG tensor. In Rb$_3$Yb(VO$_4$)$_2$, since the $^{51}$V nuclei are in a $3m.$ (axial) local symmetry, one expects $\eta=0$. Hence, $2I-1$ equally spaced ($n\nu_{\rm Q}$, with $n = 1,...,2I-1$) satellite lines corresponding to the quadrupolar resonances will appear on either side of the central transition ($I_z = +1/2 \leftrightarrow -1/2$) in $^{51}$V spectra~\cite{Lang094404}.

The temperature-dependent NMR spectra at 77.1~MHz are shown in Fig.~\ref{Fig7}. As expected for $I =7/2$, seven resonance lines are visible at higher temperatures: one central transition and three satellite pairs. Further, as the temperature is lowered, we observe asymmetric line broadening and a shift of the central line position. The absence of any abrupt line broadening at low temperatures down to 1.6~K excludes the onset of magnetic LRO~\cite{Ambika015803}. To simulate the anisotropic powder spectra, we used the isotropic ($K_{\text{iso}}$) and axial ($K_{\text{ax}}$) components of the NMR shift ($K$) defined by~\cite{slichter2013}
\begin{equation}
K = K_{\text{iso}} + K_{\text{ax}}(3\cos^2\theta - 1),
\label{eq:NMR}
\end{equation}
where $\theta$ defines the orientation of the external field relative to the hyperfine principal axis. The simulated spectrum at $T = 250$~K (inset Fig.~\ref{Fig7}) gives $K_{\rm iso} \simeq 0.036$\%, $K_{\rm ax} \simeq -0.012$\%, $\eta=0$, and $\nu_{\rm Q}\approx 0.105$~MHz. The quadrupole frequency was found to remain constant down to 1.6~K, ruling out any possible structural distortion in the compound.

\begin{figure}
\includegraphics[width=\columnwidth]{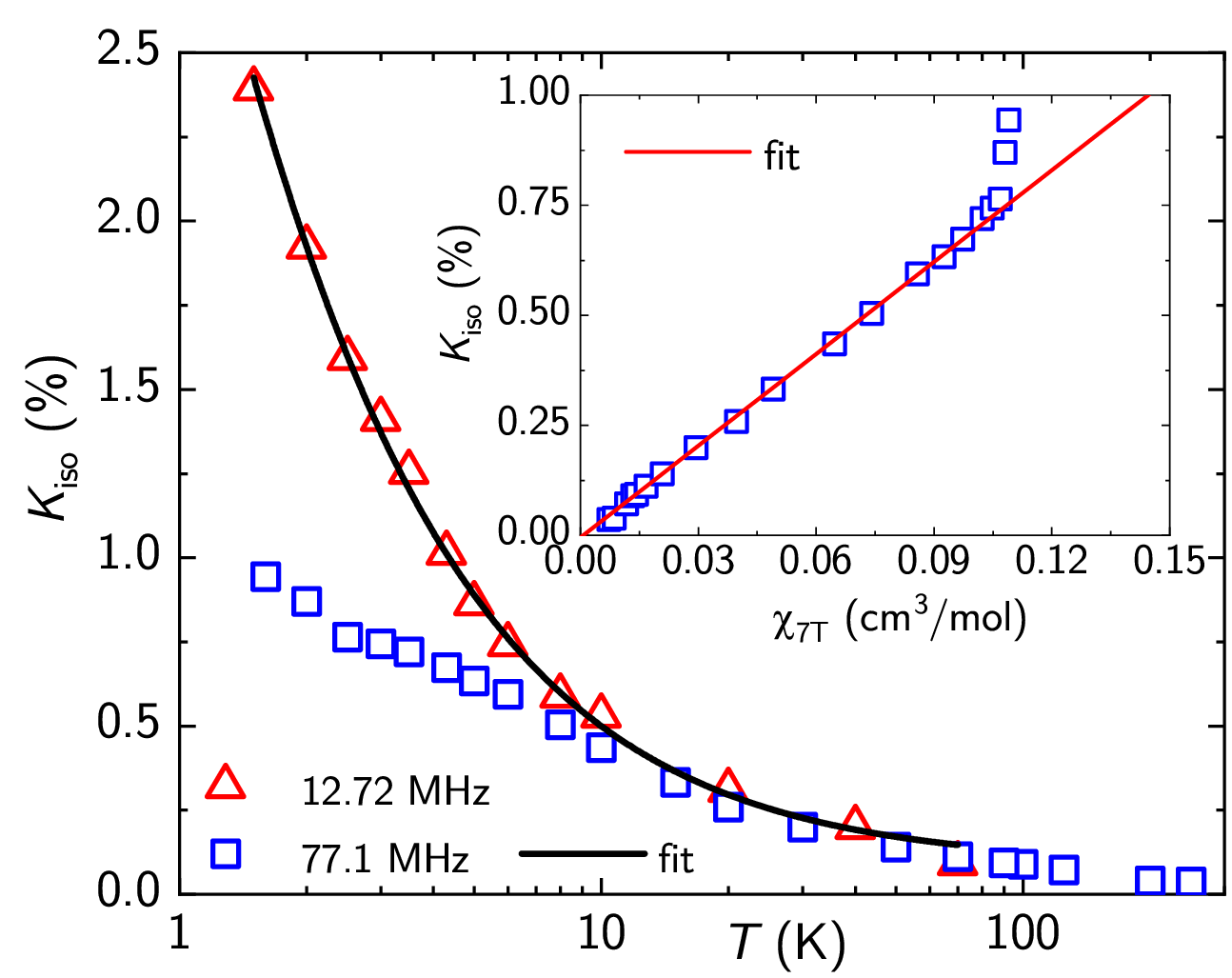}
\caption{\label{Fig8} Temperature dependence of $K_{\rm iso}$ for $^{51}$V measured at 12.72 MHz ($\sim 1.136$T) and 77.1 MHz ($\sim 6.89$T). The solid line represents the fit to the 77.1~MHz data, as discussed in the text. Inset: $K_{\rm iso}$ (77.1~MHz) vs $\chi$ measured at 7~T with the linear fit shown as solid line.}
\end{figure}
Temperature-dependent NMR shift $K(T)$ at 12.72~MHz and 77.1~MHz is shown in Fig.~\ref{Fig8}. Since $K(T)$ directly reflects the spin susceptibility $\chi_{\rm spin}$, the linear relation can be expressed as
\begin{equation}
K(T) = K_{0}+\frac{\mathcal{A}_{\rm hf}}{N_{\rm A}\mu_{\rm B}}\chi_{\rm spin},
\label{NMR_kchi}
\end{equation}
where $K_0$ is the temperature-independent orbital contribution and $\mathcal{A}_{\rm hf}$ the hyperfine coupling constant between $^{51}$V nuclear spin with the Yb$^{3+}$ electronic spins. From the linear $K$ vs $\chi$ plot (inset Fig.~\ref{Fig8}), we obtained $K_0 \simeq -0.005$\% and $\mathcal{A}_{\rm hf} \approx 0.040(1)$~T/$\mu_{\rm B}$. To estimate the exchange coupling, the data below 100~K were fitted by Eq.~\eqref{NMR_kchi}, taking $\chi_{\rm spin}$ for an isotropic $S=1/2$ TLAF model [Eq.~\eqref{HTSE}]. The fit yields $J/k_{\rm B}\approx 0.27(1)$~K and $g \simeq 2.59(1)$. While fitting the data, the hyperfine coupling was kept fixed to $0.040$~T/$\mu_{\rm B}$. This value of $J/k_{\rm B}$ is consistent with the one obtained from the $\chi(T)$ analysis.

\begin{figure}
\includegraphics[width=\linewidth, height=6.5cm]{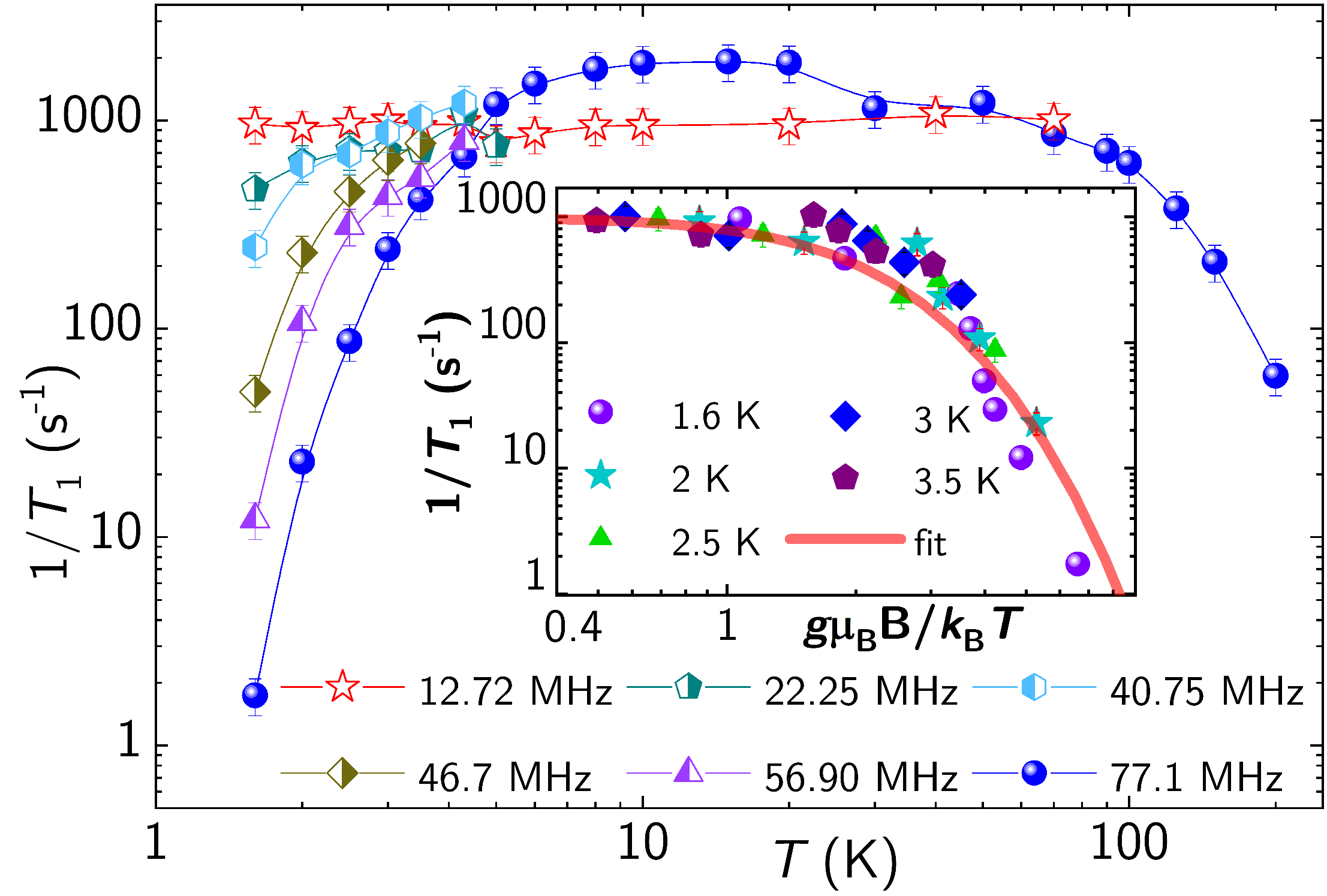}
\caption{\label{Fig9} Temperature dependence of $1/T_{1}$ at various frequencies. Inset: $1/T_1$ versus the scaling variable $g\mu_{\rm B}B/k_{\rm B}T$. The solid line represents the derivative of the Brillouin function for $S=1/2$ with $g=2.3$.}
\end{figure}
To study the spin dynamics, the spin-lattice relaxation rate ($1/T_{1}$) was obtained by selectively irradiating the central peak of the $^{51}$V NMR spectra. Usually, $1/T_{1}$ is obtained by fitting the recovery of the longitudinal magnetization [$M_z(t)$] to an exponential function appropriate for a quadrupolar nuclei with $I=7/2$~\cite{Gordon783, Sebastian064413}. However, our measured recovery curves display a broad distribution of relaxation times. Therefore, we have defined $T_1$ as the time required for $M_z(t)$ to recover to 63.2\% (i.e. $1-1/e$) of its equilibrium magnetization ($M_{\infty}$). The resulting temperature dependence of $1/T_1$ is presented in Fig.~\ref{Fig9}. Upon cooling, $1/T_1$ measured at 77.1~MHz gradually increases and exhibits a plateau below about 80~K, distinctly different from the constant behavior typical of a purely paramagnetic system~\cite{Moriya23}. The observed increase can be ascribed to the CEF effect of Yb$^{3+}$~\cite{Zeng045149}. In the fast-fluctuation limit, the slowing down of spin fluctuations follows $1/T_1 \propto \nu^{-1}$ (where, $\nu$ is the fluctuation frequency), leading to the enhancement in $1/T_1$. At low temperatures (below 10~K), $1/T_1$ shows a significant field dependence. It decreases by almost three orders of magnitude from the low field (12.72~MHz, $\sim 1.13$~T) to high field (77.1~MHz, $\sim 6.8$~T) value at 1.6~K, a signature of paramagnetic fluctuations.

To explain the characteristic behavior of $1/T_1$ at low temperatures, we followed the paramagnetic spin fluctuation model reported in Ref.~\cite{Furukawa8635}. $1/T_1$ is generally expressed through the transverse spin-spin correlation function within the Kubo linear response formalism as
\begin{equation}
\frac{1}{T_1} \propto \int_{-\infty}^{\infty} dt \, e^{i \omega_0 t} \langle S_+(t) S_-(0) \rangle,
\end{equation}
where $\omega_0$ is the nuclear Larmor frequency and $S_\pm=S_x\pm iS_y$ represent the transverse electron spin operators. In the low-frequency limit ($\omega_0\to 0$), the spin-lattice relaxation rate can be connected to the imaginary part of the dynamical susceptibility $\chi^{\perp}(\omega)$ via the fluctuation-dissipation (FD) theorem~\cite{Kubo255} as
\begin{equation}
\frac{1}{T_1} \propto \lim_{\omega \to 0} \frac{1}{\beta\omega} \, \text{Im} \chi^{\perp}(\omega).
\end{equation}
The FD theorem connects $\text{Im} \chi^{\perp}(\omega)$ to the Fourier transform of the spin-spin correlation function as: $\text{Im} \chi^{\perp}(\omega) \propto (1 - e^{-\beta \omega}) S(\omega)$, where $\beta = 1/k_B T$ and $S(\omega)$ is the power spectrum of spin fluctuations. In the low-frequency limit $\omega \to 0$, we get
\begin{equation}
\frac{1}{T_1} \propto S(\omega \to 0) \propto \int_{-\infty}^{\infty} dt \, \langle S_+(t) S_-(0) \rangle \propto \langle \delta S_z^{2}  \rangle.
\end{equation}
Here, $\sqrt{\langle \delta S_z^{2}} \rangle$ is the standard deviation expected for a perturbed Hamiltonian where the spins point along the $z$-direction with applied field $B$. In such a scenario, we can expect the average magnetization along $z$ direction as 
\begin{equation}
    \langle S_z \rangle = \frac{1}{Z} \text{Tr} [S_z \{\exp-\beta{(\mathcal{H}-BS_z})\}].
\end{equation}
We can then define the static susceptibility as
\begin{equation}
\chi=\dfrac{d\langle S_z \rangle}{dB} =\beta[\langle S^2_z \rangle - (\langle S_z \rangle)^2] = \beta\langle \delta S_z^{2} \rangle.
\end{equation}
Thus, we can relate $1/T_1$ to the field-dependent differential susceptibility as
\begin{equation}
    \frac{1}{T_1(T,B)} \propto T \frac{\partial \langle S_z \rangle}{\partial B} \propto \frac{dB_{1/2}(x)}{dx},
    \label{NMR_BF}
\end{equation}
where $B_{1/2}$ denotes the Brillouin function for $J=1/2$, and $x \equiv g\mu_{\rm B}B/(2k_{\rm B} T)$. To illustrate this relationship clearly, we plotted $1/T_1$ against the scaled variable $2x$, using a single proportionality constant in the inset of Fig.~\ref{Fig9}. The experimental data agree well with the theoretical expectation expressed in Eq.~\eqref{NMR_BF} derived in Ref.~\cite{Furukawa8635}. The consistency of this paramagnetic-like behavior observed in our NMR measurements down to 1.6~K aligns well with bulk and thermodynamic properties. To observe the deviation from this paramagnetic scenario, further measurements at significantly lower temperatures would be required, where the magnetic correlation sets in.

\section{CEF analysis}
To account for the influence of local ligand environment on the magnetic properties of Rb$_3$Yb(VO$_4$)$_2$, we performed a crystal electric field (CEF) analysis based on the point charge approximation. The CEF Hamiltonian is given by~\cite{Stevens209}
\begin{equation}\label{CEF}
\mathcal{H}_{\rm CEF} = \sum_{l,m}B_l^m\hat{O}_l^m.
\end{equation}
Here, $\hat{O}_l^m$ are the Stevens operators, representing angular momentum operators projected onto the site symmetry ($D_{3d}$) of the Yb$^{3+}$ ions in Rb$_3$Yb(VO$_4$)$_2$~\cite{Hutchings227,Stevens209}. This formulation considers the electrostatic potential generated by ligands treated as point charges. The CEF parameters $B_l^m$ characterize the strength and symmetry of this potential and are determined by the spatial distribution of ligands around the magnetic ion. For a $f$-electron system, the index $l$ takes even values from 0 to 6, while $m$ ranges from $-l$ to $l$.

\begin{figure}
\includegraphics[scale=0.3]{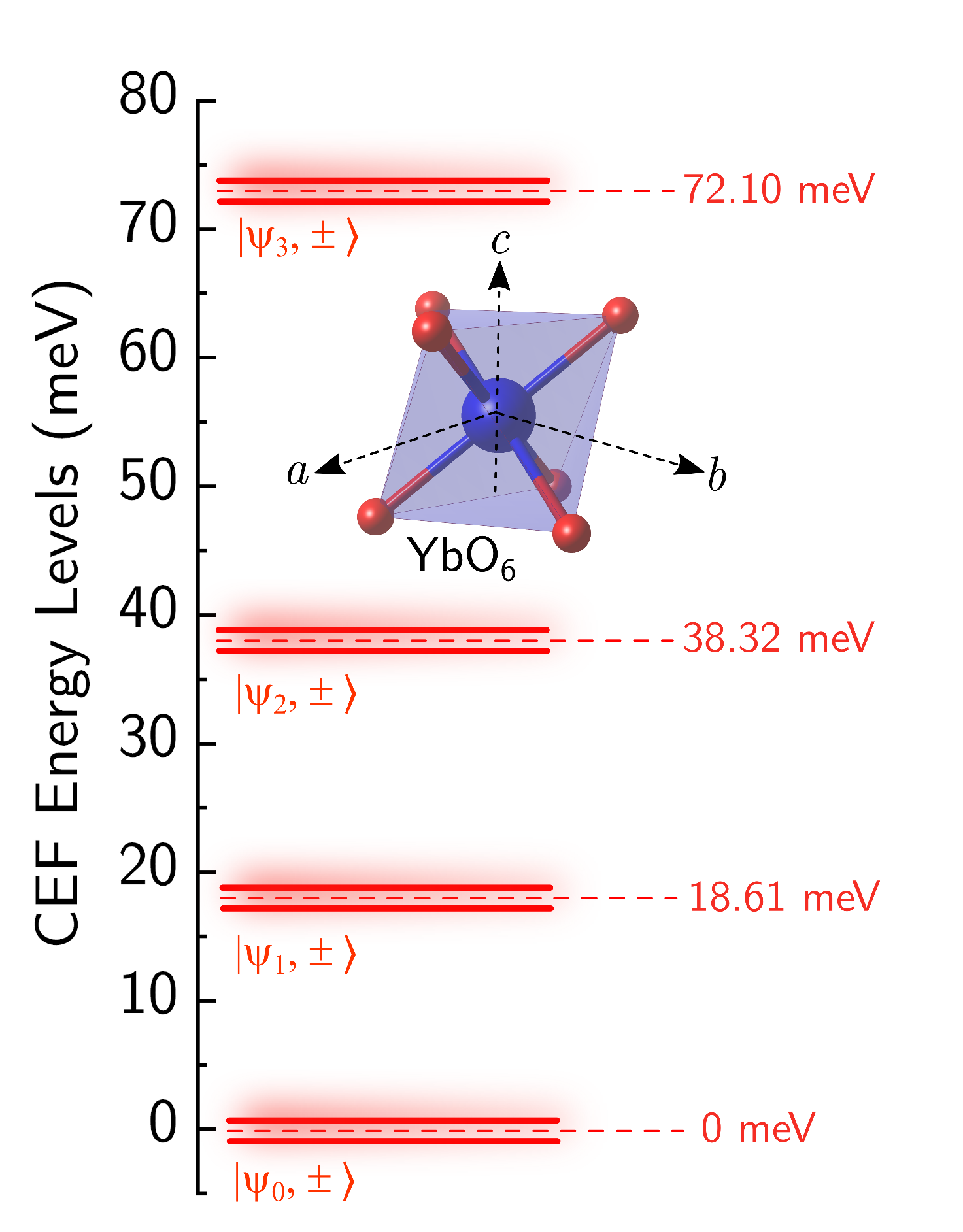}
\caption{Schematic representation of the CEF energy levels ($J_z =\pm1/2$, $\pm3/2$, $\pm5/2$, and $\pm7/2$) in Rb$_3$Yb(VO$_4$)$_2$ derived from point charge model calculations. The local coordination environment of Yb$^{3+}$ is illustrated by regular YbO$_6$ octahedra formed by surrounding O$^{2-}$ ions, which generate the CEF potential at the Yb$^{3+}$ site.}
\label{Fig10}
\end{figure}
We computed the $B_l^m$ coefficients using the \verb|PyCrystalField| python package~\cite{Hutchings227} and the obtained non-zero CEF parameters for Rb$_3$Yb(VO$_4$)$_2$ are summarized in Table~\ref{CEF_Para}.
\begin{table}[h]
    \centering
    \setlength{\tabcolsep}{0.3cm}
    \renewcommand{\arraystretch}{1.3}
    \caption{Calculated CEF parameters for Rb$_3$Yb(VO$_4$)$_2$.}
    \label{CEF_Para}
    \begin{tabular}{|c|c|}
        \hline
        \textbf{$B_l^m$} & \textbf{Values (meV)} \\
        \hline
        $B_2^0$ & $-8.925\times10^{-1}$ \\
        $B_2^1$ & $-1.00\times10^{-8}$  \\
        $B_4^0$ & $2.1\times10^{-2}$ \\
        $B_4^3$ & $6.647\times10^{-1}$ \\
        $B_6^0$ & $1.596\times10^{-4}$ \\
        $B_6^3$ & $-7.016\times10^{-4}$ \\
        $B_6^6$ & $1.44\times10^{-3}$ \\
        \hline
    \end{tabular}
\end{table}
Using the calculated $B_l^m$ parameters, we constructed the CEF Hamiltonian and obtained its energy eigenvalues by numerical diagonalization. The resulting CEF level scheme consists of four Kramers' doublets located at 0, 18.61, 38.32, and 72.10~meV, respectively (see Fig.~\ref{Fig10}). A well defined energy separation between the lowest and first excited doublets ($\sim 210$~K) confirms that the ground state is clearly governed by a $J_{\rm eff} = 1/2$ Kramers' doublet~\cite{Guchhait2025}.

The eigenfunctions of these doublets can be expressed as a linear combination of the $|J, m_J\rangle$ basis states
\begin{equation}
\label{CEF_wave_vector}
|\psi_k,\pm\rangle = \sum_{m_J = -\frac{7}{2}}^{\frac{7}{2}} C_{m_J}^{k,\pm} \left|J = \frac{7}{2}, m_J \right\rangle,
\end{equation}
where $C_{m_J}^{k,\pm}$ are the expansion coefficients and $k = 0, 1, 2, 3$ index the CEF levels. The composition of the ground state doublet ($k = 0$) for Rb$_3$Yb(VO$_4$)$_2$ is found to be
\begin{equation}
\label{Wavefunction}
|\psi_0,\pm\rangle = - 0.732\left|\mp\frac{5}{2}\right\rangle \mp 0.451\left|\pm\frac{1}{2}\right\rangle + 0.511\left|\pm\frac{7}{2}\right\rangle,
\end{equation}
highlighting significant admixture of multiple $m_J$ components. The energies and corresponding eigenvector coefficients for all Kramers' doublets are summarized in Table~\ref{Eigenvalue_and_Eigervector}. As one can see from Eq.~\eqref{Wavefunction}, in the ground state eigenfunction, there are comparable contributions from $\left|\pm\frac{1}{2}\right\rangle$, $\left|\pm\frac{5}{2}\right\rangle$, and $\left|\pm\frac{7}{2}\right\rangle$ states. While a large weightage of $\left|\pm\frac{1}{2}\right\rangle$ indicates quantum effects/tunneling between states, the contributions from $\left|\pm\frac{5}{2}\right\rangle$ and $\left|\pm\frac{7}{2}\right\rangle$ also reflects equally probable classical nature. Similar observations are reported in several rare-earth based systems~\cite{Scheie144432,Guchhait144434}.

\begin{table*}
\caption{Energy eigenvalues and the weightage factors ($C_{m_J}^{k,\pm}$) corresponding to four Kramers' doublets for Rb$_3$Yb(VO$_4$)$_2$.}
\label{Eigenvalue_and_Eigervector}
\begin{ruledtabular}
\begin{tabular}{c|cccccccccc}
$E$ (meV) & $|-\frac{7}{2}\rangle$ & $|-\frac{5}{2}\rangle$ & $|-\frac{3}{2}\rangle$ & $|-\frac{1}{2}\rangle$ & $|\frac{1}{2}\rangle$ & $|\frac{3}{2}\rangle$ & $|\frac{5}{2}\rangle$ & $|\frac{7}{2}\rangle$ \tabularnewline
 \hline 
0.00 & 0 & -0.732 & 0 & 0 & -0.451 & 0 & 0 & 0.511 \tabularnewline
0.00 & 0.511 & 0 & 0 & 0.451 & 0 & 0 & -0.732 & 0 \tabularnewline
18.61 & 0 & 0.631 & 0 & 0 & -0.165 & 0 & 0 & 0.758 \tabularnewline
18.61 & -0.758 & 0 & 0 & -0.165 & 0 & 0 & -0.631 & 0 \tabularnewline
38.32 & 0 & 0 & -0.988 & 0 & 0 & -0.154 & 0 & 0 \tabularnewline
38.32 & 0 & 0 & 0.154 & 0 & 0 & -0.988 & 0 & 0 \tabularnewline
72.10 & 0 & 0.258 & 0 & 0 & -0.877 & 0 & 0 & -0.405  \tabularnewline
72.10 & -0.405 & 0 & 0 & 0.877 & 0 & 0 & 0.258 & 0 \tabularnewline
\end{tabular}\end{ruledtabular}
\end{table*}

\begin{figure*}
\includegraphics[width=\linewidth]{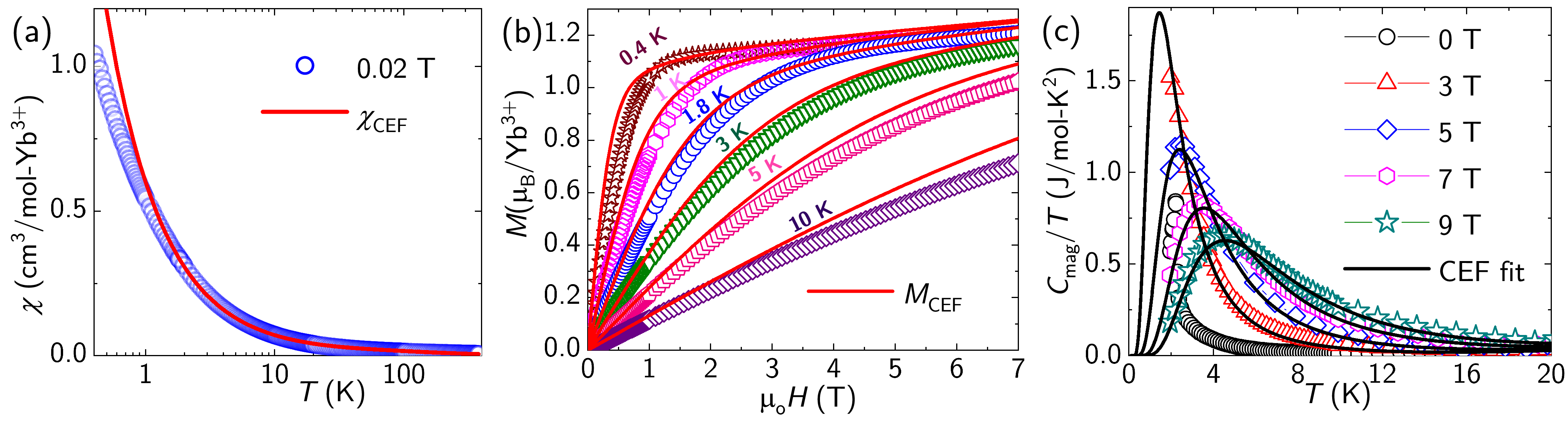}
\caption{ (a) Comparison of the experimental $\chi(T)$ of Rb$_3$Yb(VO$_4$)$_2$ with the susceptibility calculated using the CEF model. (b) Isothermal magnetization curves simulated at various temperatures based on the CEF model, shown alongside the corresponding experimental data. (c) Calculated CEF specific heat [$C_{\rm CEF}$ vs $T$] at different magnetic fields, compared with the experimental $C_{\rm mag}$ data.}
\label{Fig11}
\end{figure*}
By utilizing the CEF parameters obtained from the point charge model, we simulated the magnetic susceptibility and magnetization of Rb$_3$Yb(VO$_4$)$_2$. The magnetization along a given direction $\alpha$ is calculated using the relation
\begin{equation}
M_{\alpha} = N_{\rm A} g \mu_{\rm B} \underbrace{\left\langle \frac{1}{Z} \sum_k e^{-\frac{E_k(H)}{k_{\rm B}T}} \langle \psi_k(H) | \hat{J}_{\alpha} | \psi_k(H) \rangle \right\rangle}_{\langle J_{\alpha} \rangle}.
\end{equation}
Here, $|\psi_k(H)\rangle$ are the eigenstates of the effective Hamiltonian $\mathcal{H} = \mathcal{H}_{\rm CEF} + g\mu_{\rm B} \textbf{B} \cdot \textbf{J}$, with $\textbf{B}$ representing the applied magnetic field vector. The partition function is given by $Z = \sum_k e^{-E_k(H)/k_{\rm B}T}$, summing over all CEF-split states. The magnetic susceptibility ($\chi_{\rm CEF}$) is obtained by differentiating $M_{\alpha}$ with respect to the applied field. As shown in Figs.~\ref{Fig11}(a,b), the calculated CEF contributions compare well with the experimental magnetic susceptibility and magnetization isotherm data.

Furthermore, we calculated the crystal field contribution to the specific heat ($C_{\rm CEF}$) by considering thermal excitations between all Zeeman-split CEF levels. For an $N$-level system, $C_{\rm CEF}$ can be expressed as
\begin{equation}
C_{\rm CEF}(T, H) = \frac{R}{(Zk_{\rm B}T)^2} \sum_{n>m}^{N} [\Delta E_{n,m}(H)]^2 e^{-\frac{E_n + E_m}{k_{\rm B}T}}.
\label{C_CEF}
\end{equation}
Here, $\Delta E_{n,m} = E_n - E_m$ is the energy difference between the $n^{\rm th}$ and $m^{\rm th}$ CEF levels. The calculated $C_{\rm CEF}$ successfully reproduces the broad maximum observed in the low-temperature specific heat, as shown in Fig.~\ref{Fig11}(c). This Schottky-type anomaly indicates that the low-temperature thermodynamic properties are predominantly governed by the Zeeman-split ground state doublet, while contributions from higher excited states are negligible. This further supports the realization of a $J_{\rm eff} = 1/2$ ground state at low temperatures.

The magnetic anisotropy of the ground state was also evaluated through the $g$-tensor components derived from the CEF analysis. The obtained values, $g_{\perp} \simeq 3.19$ and $g_{z} \simeq 0.74$, result in an average $g$-value of $g_{\rm avg} = (2g_{\perp} + g_{z})/3 \simeq 2.37$, which matches well with the value obtained from the magnetization and specific heat analysis. The pronounced anisotropy, with $g_{\perp} \gg g_{z}$, reflects the dominant in-plane character of the magnetic moments, consistent with the layered triangular lattice geometry of Rb$_3$Yb(VO$_4$)$_2$.

\section{Summary}
In summary, we have investigated the structural, static, and dynamic properties of a Yb$^{3+}$-based disorder-free TLAF Rb$_3$Yb(VO$_4$)$_2$ using different measurement techniques, upheld by CEF analysis. Our powder XRD confirms the hexagonal structure with no signature of site disorder, essential for probing intrinsic quantum phenomena in TLAFs. The ground state is found to be a Kramers' doublet with $J_{\rm eff} = 1/2$. Magnetic susceptibility and magnetization measurements indicate weak AFM correlation ($\theta_{\rm CW}^{\rm LT}\simeq -0.26$~K) among the $J_{\rm eff} = 1/2$ spins. The specific heat data analysis using a non-magnetic analog demonstrates that at low-temperatures and in different magnetic fields it is dominated by a Schottky anomaly arising from the Zeeman split Kramers' doublets. An average exchange coupling of $J/k_{\rm B} \simeq 0.18$~K is consistently reproduced from the analysis of $\theta_{\rm CW}^{\rm LT}$ and HTSE fit to the low-$T$ $\chi(T)$ data. The $^{51}$V NMR spectra feature quadrupole effect as expected for an $I=7/2$ nucleus. The absence of magnetic LRO down to 1.6~K was confirmed from both NMR spectral and $1/T_1$ measurements. $1/T_1$ that probes the dynamic susceptibility reveals significant paramagnetic fluctuations at low temperatures, consistent with the bulk magnetic measurements. From the $K$ vs $\chi$ plot, the hyperfine coupling between the $^{51}$V nucleus and Yb$^{3+}$ spins is estimated to be $A_{\rm hf} \simeq 0.04(1)$~T/$\mu_{\rm B}$. Using the point charge CEF calculations, the energy eigenvalues are estimated which are used to replicate the experimental thermodynamic data.
These comprehensive results emphasize the potential of Rb$_3$Yb(VO$_4$)$_2$ as a disorder-free system for exploring quantum fluctuations and possible QSL behavior in Yb-based TLAFs at low temperatures.

\textit{Note added}: During the course of our work, we became aware of the work by Ma $et.~al$.~\cite{Ma155141} where the authors have reported magnetization, specific heat, and inelastic neutron scattering experiments. They found no evidence for magnetic LRO down to 100~mK and proposed a non-trivial disordered ground state for Rb$_3$Yb(VO$_4$)$_2$.

\section{acknowledgments}
S.J.S acknowledges Fulbright-Nehru Doctoral Research Fellowship Award No.~2997/FNDR/2024-2025 and the Prime Minister's Research Fellowship (PMRF) scheme, Government of India to be a visiting research scholar at the Ames National Laboratory. S.J.S, R.K, A.B, and R.N acknowledge SERB, India, for financial support bearing sanction Grant No.~CRG/2022/000997 and DST-FIST with Grant No.~SR/FST/PS-II/2018/54(C). The research was supported by the U.S. Department of Energy, Office of Basic Energy Sciences, Division of Materials Sciences and Engineering. Ames National Laboratory is operated for the U.S. Department of Energy by Iowa State University under Contract No.~DEAC02-07CH11358.
\appendix
\section{}
\label{appx}
Consider a quantum two-level system with eigenstates $\ket{\phi}$ and $\ket{\psi}$ Zeeman split levels with an energy gap $\Delta$ as shown in the inset of Fig.~\ref{Fig4}. The energies of these two states are $E_1 = -\Delta/2$ and $E_2 = \Delta/2$. Let the magnetic moment operator $\mu$ has matrix elements between the eigen states $\ket{\phi}$ and $\ket{\psi}$ given by
\[
\mu =
\begin{pmatrix}
	\langle \phi | \mu | \phi \rangle & \langle \phi | \mu | \psi \rangle \\
	\langle \psi | \mu | \phi \rangle & \langle \psi | \mu | \psi \rangle
\end{pmatrix}
=
\begin{pmatrix}
	S_\phi & S_{12} \\
	S_{12}^* & S_\psi 
\end{pmatrix}.
\]
In the weak-field limit, one can define the magnetic susceptibility ($\chi$) as
\begin{equation}
\label{chi_fullexp}
\chi = \frac{N}{k_{\rm B} T} \langle \mu^2 \rangle = \frac{N}{k_{\rm B} T} \left[ \langle \mu^2 \rangle_{\text{diag}} + \langle \mu^2 \rangle_{\text{off-diag}} \right].
\end{equation}
The diagonal (Curie-like) contribution arises from the mean square thermal average of the magnetic moment as
\begin{equation}
\langle \mu^2 \rangle_{\text{diag}} = \sum_i \langle i | \mu | i \rangle^2 \cdot \frac{e^{-\beta E_i}}{Z}.
\label{Sus}
\end{equation}
Here, $Z$ is the partition function and for a two-level system it is simply $Z= 2 \cosh\left( \frac{\beta \Delta}{2} \right)$ and $\beta=1/k_{\rm B}T$.
Thus, for the diagonal part, we can write
\begin{equation}
	\langle \mu^2 \rangle_{\text{diag}} = S_\phi^2 \cdot \frac{e^{\beta \Delta/2}}{Z} + S_\psi^2 \cdot \frac{e^{-\beta \Delta/2}}{Z}.
\end{equation}
The off-diagonal (Van-Vleck) contribution comes from virtual transitions between the states $\ket{\phi}$ and $\ket{\psi}$ which is given by
\begin{equation}
\langle \mu^2 \rangle_{\text{off-diag}} = \sum_{i \ne j} \frac{|\bra{i} \mu \ket{j}|^2}{E_j - E_i} \cdot \frac{e^{-\beta E_i} - e^{-\beta E_j}}{Z}.
\end{equation}
Thus, for a two-level system, the mean squared of the off-diagonal terms of the magnetic moment can be written as
\begin{equation}
	\langle \mu^2 \rangle_{\text{off-diag}} = \frac{S_{12}^2}{\Delta} \cdot \frac{e^{\beta \Delta/2} - e^{-\beta \Delta/2}}{Z}
	= \frac{S_{12}^2}{\Delta} \cdot \tanh\left( \frac{\beta \Delta}{2}
    \right).
\end{equation}
Adding these two contributions, Eq.~\eqref{chi_fullexp} takes the form
\begin{equation}
	\begin{split}
		\chi = \frac{N}{k_B T} \Bigg[ & S_\phi^2 \cdot \frac{e^{\beta \Delta/2}}{Z}
		+ S_\psi^2 \cdot \frac{e^{-\beta \Delta/2}}{Z} \\
		& + S_{12}^2 \cdot \left( \frac{1}{\Delta} \right)
		\tanh\left( \frac{\Delta}{2 k_B T} \right) \Bigg].
	\end{split}
\end{equation}
We define the coefficients; $C_0 = \frac{1}{2}(S_\phi^2 + S_\psi^2), C_1 = S_{12}^2$, and $C_2 = \frac{1}{2}(S_\phi^2 - S_\psi^2)$. The total susceptibility becomes
\begin{equation}
\begin{split}
\chi = \frac{N}{k_B T} \Bigg[ 
		& C_0 
		+  C_1\left( \frac{2k_B T}{\Delta} \right) 
		\tanh\left( \frac{\Delta}{2k_B T} \right) \\
		& + C_2 \tanh\left( \frac{\Delta}{2k_B T} \right) 
		\Bigg].
	\end{split}
    \label{VV_KD}
\end{equation}
This result includes both Curie-like and Van-Vleck contributions in a unified framework for the Kramers' doublet. For fitting the experimental $\chi T$ data, we used the modified form of Eq.~\eqref{VV_KD}
\begin{equation}
\begin{split}
\chi = \chi_0 +\frac{N}{k_B (T-T_0)} \Bigg[ 
		& C_0 
		+  C_1\left( \frac{2k_B T}{\Delta} \right) 
		\tanh\left( \frac{\Delta}{2k_B T} \right) \\
		& + C_2 \tanh\left( \frac{\Delta}{2k_B T} \right) 
		\Bigg].
	\end{split}
    \label{VV_KD_appx}
\end{equation}
Here, $T_0$ represents the energy scale of $\theta_{\rm CW}$.


%
	
\end{document}